\documentclass[twocolumn]{aa} 

\usepackage{graphicx}
\usepackage{txfonts}
\usepackage{natbib} 
\bibpunct{(}{)}{;}{a}{}{,} 
\begin{document}
   \title{Extinction controlled adaptive phase-mask coronagraph}

   \author{P.~Bourget, N.~Schuhler, D.~Mawet, P.~Haguenauer }

   \institute{European Southern Observatory, Alonso de Cord\'ova 3107, Vitacura, Santiago, Chile\\
              \email{pbourget@eso.org}}

   \date{Accepted October 26, 2012}

 
  \abstract
  {Phase-mask coronagraphy is advantageous in terms of inner working angle and discovery space. It is however still plagued by drawbacks such as sensitivity to tip-tilt errors and chromatism. A nulling stellar coronagraph based on the adaptive phase-mask concept using polarization interferometry is presented in this paper.}
  {Our concept aims at dynamically and achromatically optimizing the nulling efficiency of the coronagraph, making it more immune to fast low-order aberrations (tip-tilt errors, focus, ...).}
  {We performed numerical simulations to demonstrate the value of the proposed method. The active control system will correct for the detrimental effects of image instabilities on the destructive interference. The mask adaptability both in size, phase and amplitude also compensates for manufacturing errors of the mask itself, and potentially for chromatic effects. Liquid-crystal properties are used to provide variable transmission of an annulus around the phase mask, but also to achieve the achromatic $\pi$ phase shift in the core of the PSF by rotating the polarization by $180^\circ$.}
  {We developed a new concept and showed its practical advantages using numerical simulations. This new adaptive implementation of the phase-mask coronagraph could advantageously be used on current and next-generation adaptive optics systems, enabling small inner working angles without compromising contrast.}
  {}
  
   \keywords{Instrumentation: high angular resolution --
   Instrumentation: adaptive optics --
                Techniques: high angular resolution
               }

   \maketitle
%

\section{Introduction}

 High contrast imaging of extra-solar planets and close environments of bright astrophysical objects in general, such as stars or active galactic nuclei, is a challenging task. Coronagraphy is now recognized as one of the must-have tools to help take on this task. However, the technical challenges for coronagraphs are significant: intrinsic contrast capabilities, ability to perform over broad bandwidths, and inner working angle. Inner working angle (or IWA), is rigorously defined as the 50\% off-axis throughput point in $\lambda/d$ units ($\lambda$ is the observing wavelength and $d$ is the telescope diameter). Accessing small IWA ($\simeq 1-2 \lambda/d$) is considered as an edge because it provides substantial scientific and technical advantages \citep{Mawet2012a}. For instance, it opens up new discovery spaces in the inner regions of stellar systems, while allowing to reach out to more distant systems in young stellar associations \citep{Mawet2012b}. From a technical standpoint, small IWA allow reducing the size of the telescope or increase the observing wavelength to new regimes.

Since the first phase-mask coronagraph proposed by \citet{Roddier1997}, alternative mask concepts have been developed to decrease the chromatic dependence of the coronagraph efficiency. \citet{Rouan2000} proposed the four quadrant phase mask to circumvent the wavelength dependence of the size of the phase mask. \citet{Riaud2001} proposed the use of multiple thin films in a four quadrant scheme to realize an achromatic $\pi$ phase shift over a broad band corresponding to usual astrophysical broadband filters. \citet{Mawet2005,Mawet2006} then described the use of achromatic retarders made out of natural and artificial birefringent materials to the same end.  The vector vortex coronagraph proposed by \citet{Mawet2005} pushed further the use of the geometrical phase \citep{Pancharatnam1956, Berry1987} in phase-mask coronagraphy, offering a new tool to solve both chromatic limitations of the phase-mask coronagraph family.

Tremendous progress has thus been accomplished this past two decades on these fronts, but focusing mostly on improving coronagraph concepts and technologies, with perhaps a lack of emphasis on system optimization as a whole \citep{Guyon2005}. For instance, one of the difficulties of the task of accessing small IWA is that systems become very sensitive to low-order aberrations such as tip-tilt \citep{Guyon2005}. Once this problem was recognized, substantial efforts went into stabilizing the system feeding the coronagraph with, e.g.~so-called low-order wavefront sensors, which are now paramount \citep{Guyon2009}. 

Our original approach presented in this paper aims at integrating the IWA capability and the mitigation of sensitivity to low-order aberrations within the coronagraph itself. The \emph{adaptive phase-mask} coronagraph concept is an innovative answer to this fundamental trade-off of small IWA coronagraphy. Contrary to previous limited-scope solutions, our concept is also applicable to both low and high-Strehl regimes, corresponding to current and next-generation adaptive optics systems, respectively.

Indeed, in the low-Strehl regime (or Ex-AO on faint targets, when the AO wavefront sensor is severely photon starved), the adaptive coronagraph can dynamically compensate fast moving low-order aberrations to guarantee a modest but fairly consistent contrast level over a wide range of conditions and natural guide star magnitudes. This mode enables scientifically rich programs such as surveys for binarity, which are not too demanding in contrast capabilities, but are very time-consuming. In the high-Strehl regime, where performances are limited by slow varying low-order aberrations unseen by the main adaptive optics wavefront sensor, the idea is to morphologically adapt the coronagraph's phase and amplitude response in quasi real time to optimize contrast at the science image level.


\section{The Adaptive Phase Mask (APM)}

The concept of adaptive coronagraphy is not new. Indeed, the Adaptive Mask Coronagraph using Ònon-solidÓ masks was developed 10 years ago to improve coronagraphs observational efficiency \citep{Bourget2004,Bourget2006}. The first version of this concept was the Hg-mask Lyot coronagraph dedicated to an astrometric survey of faint satellites near Jovian planets \citep{Bourget2001,Vieira2004,Veiga2006}. To optimize the occulting process of the Lyot coronagraph, a compressed mercury (Hg) drop was used as an occulting disk, and its controllable diameter enabled true dynamical optimization to the seeing conditions, or to the required fraction of the Airy diameter. 

Our new adaptive phase-mask coronagraph is the logical evolution of these early developments, and the Roddier \& Roddier coronagraph concept \citep{Roddier1997}. Indeed, the nulling efficiency provided by a classical phase mask is strongly dependent on the amplitude balance of the two phase shifted waves. The main factors contributing to the unbalance of the amplitudes are: the wavelength dependence of the Airy disk diameter (i.e.~of the phase mask diameter), the image centering stability on the mask and the Strehl ratio. The optimal phase mask diameter is essentially wavelength and bandwidth dependent and must be controlled with high accuracy; it can also depend on the pupil apodization technique used \citep{Guyon2006}. Even using an adaptive optics correction, the residual low-order aberrations will impact the centering and the Airy pattern shape. Instead of acting only on the origin of the amplitude misbalance, we propose to compensate for it by actively balancing the amplitudes via a modulation of the transmission of the area outside of the phase mask. The control system will correct for both the inadequate phase mask diameter and the effects of image instability on the destructive interference. The control loop error signal is simply obtained by a direct measurement of the nulling efficiency with an avalanche photodiode (APD) at the coronagraph output (Fig.~\ref{fig3}). The optical modulation is done by the means of liquid-crystal polarization properties also used to produce the $\pi$ phase shift.

\subsection{Breakthrough use of the ``geometrical'' phase}

In this section, we show why the use of the ``geometrical'' Pancharatnam-Berry phase \citep{Pancharatnam1956,Berry1987} allowed us to solve the original problems of the Roddier phase mask while providing leverage to perform adaptive size and phase/amplitude optimization of its profile. Note that the ``geometrical'' phase has already been successfully demonstrated in phase-mask coronagraphy \citep{Mawet2005, Murakami2010}.

\subsubsection{The central $\pi$ phase shift}

\begin{figure}[!ht]
  \centering
\includegraphics[scale=1.1]{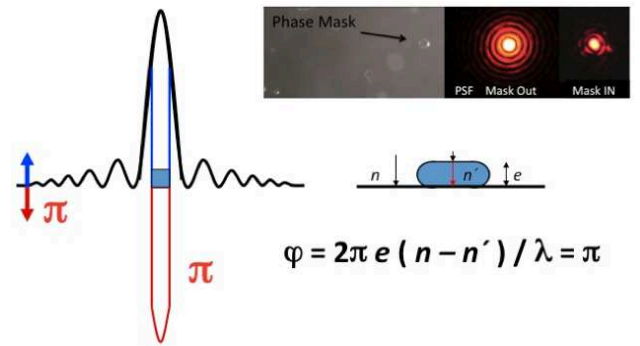}
  \caption{Early concept of the adaptive phase mask, using a gas bubble immersed in a layer of oil between two optical windows, whose thickness can be controlled to modulate the optical path difference. Owing to the coupling between diameter and thickness, this concept could only work over a narrow wavelength range. See text for details.  \label{fig1}}
\end{figure}

The first version of the Adaptive Phase Mask (APM) was made of a gas bubble immersed in a layer of oil between two optical windows (Fig.~\ref{fig1}). The phase delay of the Airy pattern core at the focal plane was obtained by refraction, following $\Delta \phi=2\pi/\lambda * (n'-n) * e$, where $\lambda$ is the observing wavelength, $n$ and $n'$ are the refractive indices of the liquid and of the bubble, respectively, and where $e$ is the distance between the optical windows of the masks. Typical values for $n$ and $n'$ are respectively about 1.5 and 1 leading to a controlled thickness $e \approx k\lambda$, where $k$ is an integer. Since the variation of the optical path delay is a linear function of $e$ and the diameter varies with $e^{-1/2}$, the thickness $e$ allows controlling either the chromatism or the diameter. As a consequence the gas bubble mask works only over a narrow wavelength range. Calibrated micro spheres inside the crystal-liquid were used to calibrate the thickness and therefore the spectral bandwidth of each adaptive phase mask. The complexity to obtain the exact bubble diameter size and the chromatism were the essentials limitations to provide high contrast efficiency during the first tests carried out at the Rio de Janeiro National Observatory -- Brasil \citep{Bourget2004,Bourget2006}. 

\begin{figure*}[!ht]
  \centering
\includegraphics[scale=2.0]{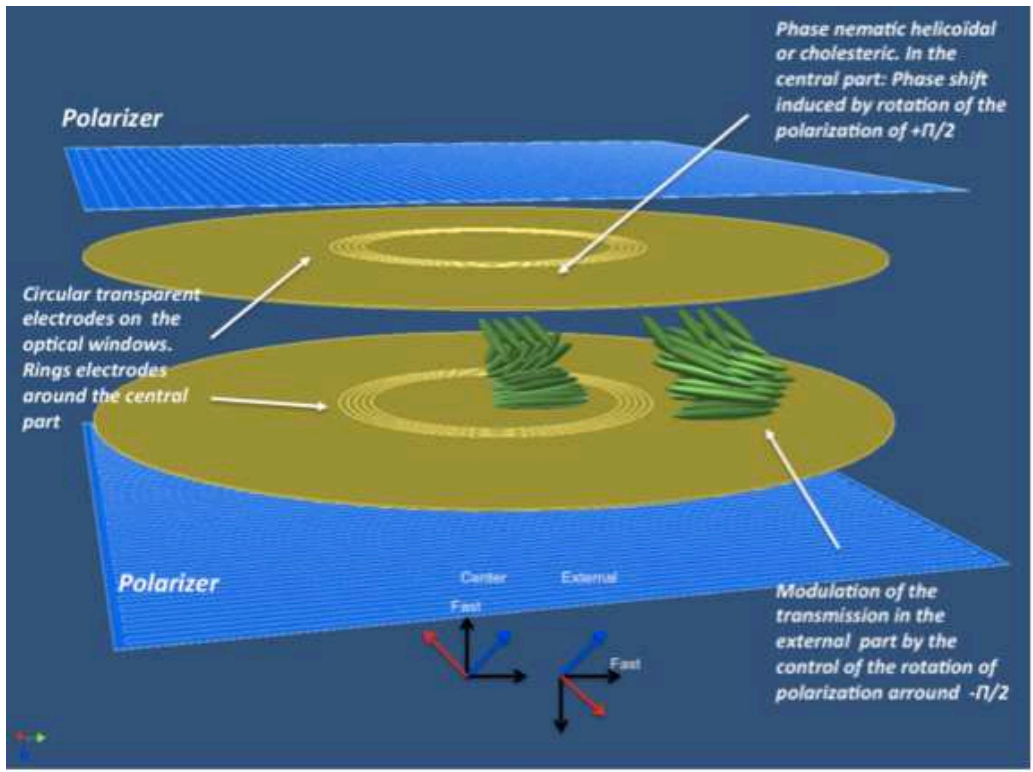}
  \caption{Improved adaptive phase-mask coronagraph concept using the ``geometrical'' or Pancharatnam-Berry phase to provide an achromatic $\pi$ phase shift between the core of the PSF and the external region. While the polarization of the core is rotated by $+\pi/2$, the outer region polarization is rotated by $-\pi/2$ creating an achromatic Rodier \& Rodier phase mask. Between the central disk electrode driving the core and the electrode driving the outer ``annulus'' a series of ring electrodes allows tuning the ratio of the surface areas phase shifted respectively by $+\pi/2$ and $-\pi/2$ . This enables the adaptation of the phase mask diameter to the observation conditions. In addition, the liquid-crystals are sandwiched between two crossed linear polarizers in order to modulate the outer region wave amplitude. This enables the fine-tuning of the amplitude balance of the two phase-shifted waves and therefore the optimization of the nulling efficiency.
   \label{fig2}}
\end{figure*}

The solution to this limitation, presented here, is based on the use of a ``geometrical'' phase shift \citep{Pancharatnam1956,Berry1987}. Instead of a phase delay induced by the refraction, a ``geometrical'' phase shift is realized by rotation of the polarization exploiting the properties of twisted nematic or cholesteric liquid-crystals. The liquid crystal is used between two perpendiculars linear polarizers (Fig.~\ref{fig2}). An achromatic $\pi$ phase shift between the core of the PSF and the external region is achieved by rotating the polarizations. The central zone is rotated by $+\pi/2$ and outside the polarization is rotated by $-\pi/2$, creating a true Roddier-type coronagraph, whose $\pi$ phase shift is intrinsically achromatic owing to the pure geometrical phase.

\subsubsection{Amplitude modulation}
In our concept, the amplitude of the wave traveling in the outer region of the mask can be modulated to optimize the balance of the interfering waves amplitudes and therefore the nulling efficiency. This is done as in any twisted-nematic amplitude modulator. The voltage applied on the outer region ``annulus'' electrode is tuned to adjust the orientation of the linear polarization state incident on the second polarizer and the amount of light transmitted. The polarizer garantees that the linear polarization coming out is effectively rotated by $-\pi/2$ although it had been rotated by a smaller amount by the liquid-crystals.
The control of the amplitude allows keeping the phase mask efficiency at a maximal level.

\subsubsection{Adaptation of the phase-mask diameter}

Built in our phase-mask concept is the capability of dynamically modulating the size of the central phase mask. 

In order to enabe the adaptability of the phase-mask diameter, we designed a transition zone, piloted by a set of concentric ring electrodes, which allow switching the orientation of the polarization between $+\pi/2$ and $-\pi/2$ over their spatial extent (Fig.~\ref{fig2}), effectively providing the choice of the size of the Roddier coronagraph's core. Note that if the phase mask diameter is decreased, the extinction control will allow observing much closer to the central star but at the cost of an increasing exposure time. Indeed, the decrease of the flux in the wave phase-shifted by the central phase mask will be compensated for by a decrease of the transmission of the outer region. This scanning ability also allows fine-tuning the nulling efficiency depending on the observing conditions, which is critical in the low-Strehl regime.

\subsubsection{Practical considerations}
Some functional aspects require a further empirical study to fine tune the concept. For instance the transition in the effect of the discontinuity in the polarization at the edges of the electrodes needs to be quantified experimentally. The polarizers and electrodes are still transmissive optics and thus will introduce some level of wavelength dependent refraction. The thickness of the liquid crystal layers should introduce very small effects but this is still to be quantified. Liquid crystal solution producing a phase delay have already been demonstrated in the frame of high contrast imaging. \citet{Baba2002} and \citet{Murakami2010} have already presented successful liquid-crystal applications for coronagraphs and especially the eight-octant phase mask (EOPoM). Even penalized by a 10 $\mu m$ gap between the eight octants, the peak extinction level of $3\times 10^{-4}$ reached by the EOPoM provides a realistic baseline for our future experiments, especially since the Adaptive Phase Mask does not present a central dead zone.

\section{Sensitivity to Tip-Tilt }\label{sect:tiptilt}
It is well known that centering errors of the Airy pattern on the Roddier phase mask produces a misbalance of the amplitude of the two phase shifted waves \citep{Guyon99}. The consequence is a degraded nulling interference in the downstream pupil. The adaptive phase mask possesses several levers to restore the balance, including the adaptive size of the central core of the mask and the adaptive transmittance of the outer region.

\subsection{Adaptive phase mask control system}
\begin{figure}[!ht]
  \centering
\includegraphics[scale=0.8]{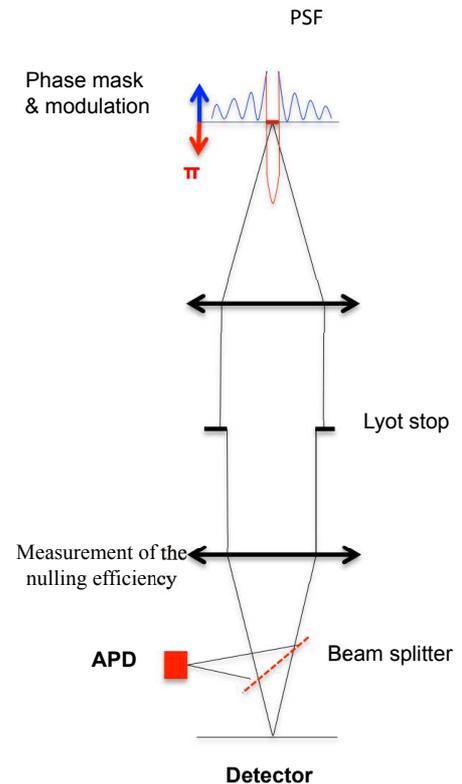}
  \caption{Adaptive Phase Mask (APM) coronagraph layout showcasing the modulation stage, Lyot stop, final science detector, and the fast and sensitive APD sensor optimally placed before the science camera and after the coronagraph/Lyot stop combo.\label{fig3}}
\end{figure}

Driving the adaptive phase mask is a simplified task. Indeed, unlike recent low-order wavefront sensors \citep{Guyon09} which aim at preventing tip-tilt excursions before they reach the occulter, and therefore require an exquisite 2-dimensional measure of the aberration, the adaptive phase-mask modifies itself in quasi real-time to provide the optimal extinction thanks to a 1-dimensional control signal. Indeed, for small tip-tilt excursions, the response of the Roddier coronagraph is a pure isotropic nulling degradation. Detecting the nulling leakage at the science camera level is all that is required to drive the amplitude compensation and/or central mask size, according to two main cases:
\begin{enumerate}
\item Low-Strehl regime: the rings close to the PSF core provide the lever arm to control the fast components of the atmospheric tip-tilt, that cannot be resolved temporally by the specific sensor due to lack of guide star photons. Such errors broaden the PSF core in a long exposure which can then be mitigated by changing the radius of the ¹ phase shifted region. The annulus shaped region, responsible for amplitude modulation, provides the lever arm to control the slow components of the atmospheric tip-tilt, that can be resolved temporally by the avalanche photodiode auxiliary detector (see below and Fig.~\ref{fig3}). Note that the bandwidth is most likely to be driven by photon noise on this detector rather than by time lags in the liquid crystal modulation device.
\item High-Strehl regime: the annulus shaped region provides the lever arm to mitigate the non-common path tip-tilt based on the auxiliary APD sensor.
\end{enumerate}

The optical setup for the adaptive mask control feedback is derived from the automatic centering and tip-tilt correction system already developed and successfully demonstrated for the Hg-Mask Lyot coronagraph at the National Observatory of Brazil \citep{Bourget2004}. The sensitivity of such a post-coronagraph sensing system is based on the high dynamic range of the pointing signal provided by the diffraction leakage induced by off-center excursions on the coronagraph. Here, we propose to use a very fast and sensitive avalanche photodiode (APD) at the output of the coronagraph. To feed the APD (Fig.~\ref{fig3}), a beam splitter is placed before the scientific detector, contrary to the differential tip-tilt sensor of SPHERE which is placed before the coronagraph \citep{Baudoz2010b}. Nominally, the flux measured by the APD will decrease and reach a minimum. This minimum will correspond to the amplitude balance of the two shifted waves and also the best nulling; this value will be used as a reference for the control loop on the transmission. While observing, if the value measured by the APD is higher than the reference one, the control loop will decrease the transmission outside of the phase mask to minimize the flux measured by the APD. As an over correction will increase the flux, the ÒnullingÓ control loop is done on a stability point. The extinction control system will maintain the nulling efficiency to an optimal value. The variance of the leakage term is therefore our objective control signal, which is minimized through a careful fine-tuning of the loop gain, which can be adapted to seeing conditions and/or adaptive optics correction residuals. Note that high band pass control is provided by the APD and Liquid-crystals properties. 

The optimization of extinction, which decreases the transmission, tends to balance the amplitudes of the phase-shifted waves. However, the destructive interference that occurs in the pupil plane is spatially dependent on the pupil energy distribution for both waves. The centering error induces a different energy repartition that will prevent the total destructive interference.

 \subsection{Quantifying the tip-tilt residuals with numerical simulations}
 
Using the Matrix Fourier Transform method presented in \citep{Soummer2007}, our numerical simulations were done to quantify the range of efficiency of the transmission control system over an increasing Tip-Tilt error.  The centering error of the Airy pattern on the adaptive phase mask was introduced in the simulation by a Tilt of the incoming wave front. The optical setup for our simulation is without central obstruction.
\begin{figure*}[!ht]
  \centering
\includegraphics[scale=1.8]{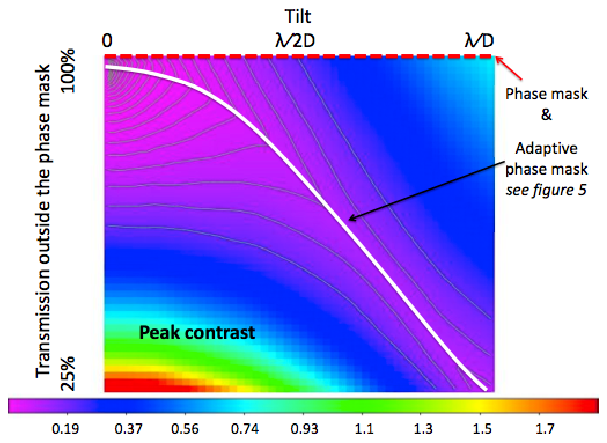}
\includegraphics[scale=1.8]{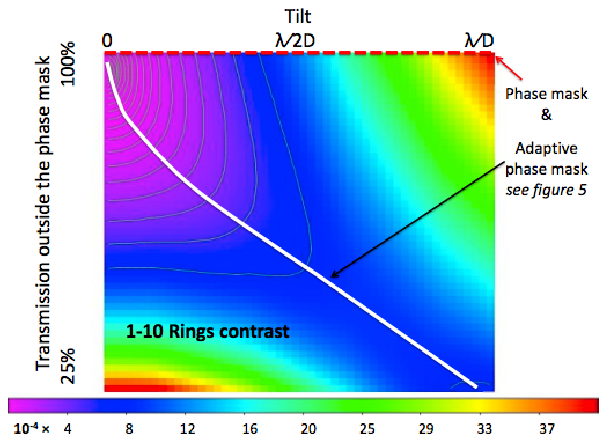}
  \caption{In this figure the contrast for the peak of the Airy pattern (top plot) and for the rings from the first until the 10th (bottom plot) is color coded for different tilt ( 0 to $\lambda/D$ over the horizontal axis) and transmission of the outer ``annulus'' region of the mask (vertical axis).  The phase mask diameter is 0.5$\lambda/D$, slightly smaller than the one that will provide a perfect theoretical balance. Numerical simulation shows that the optimal nulling efficiency of the phase mask for zero tilt is obtained for a 95\% transmission outside of the phase mask. As the tilt is increased, the control of the outer region transmission allows mitigating the degradation of the contrast as can be seen on the pink-blue valley of the color map.\label{fig4}}
\end{figure*}

\begin{figure*}[!ht]
  \centering
\includegraphics[scale=0.4]{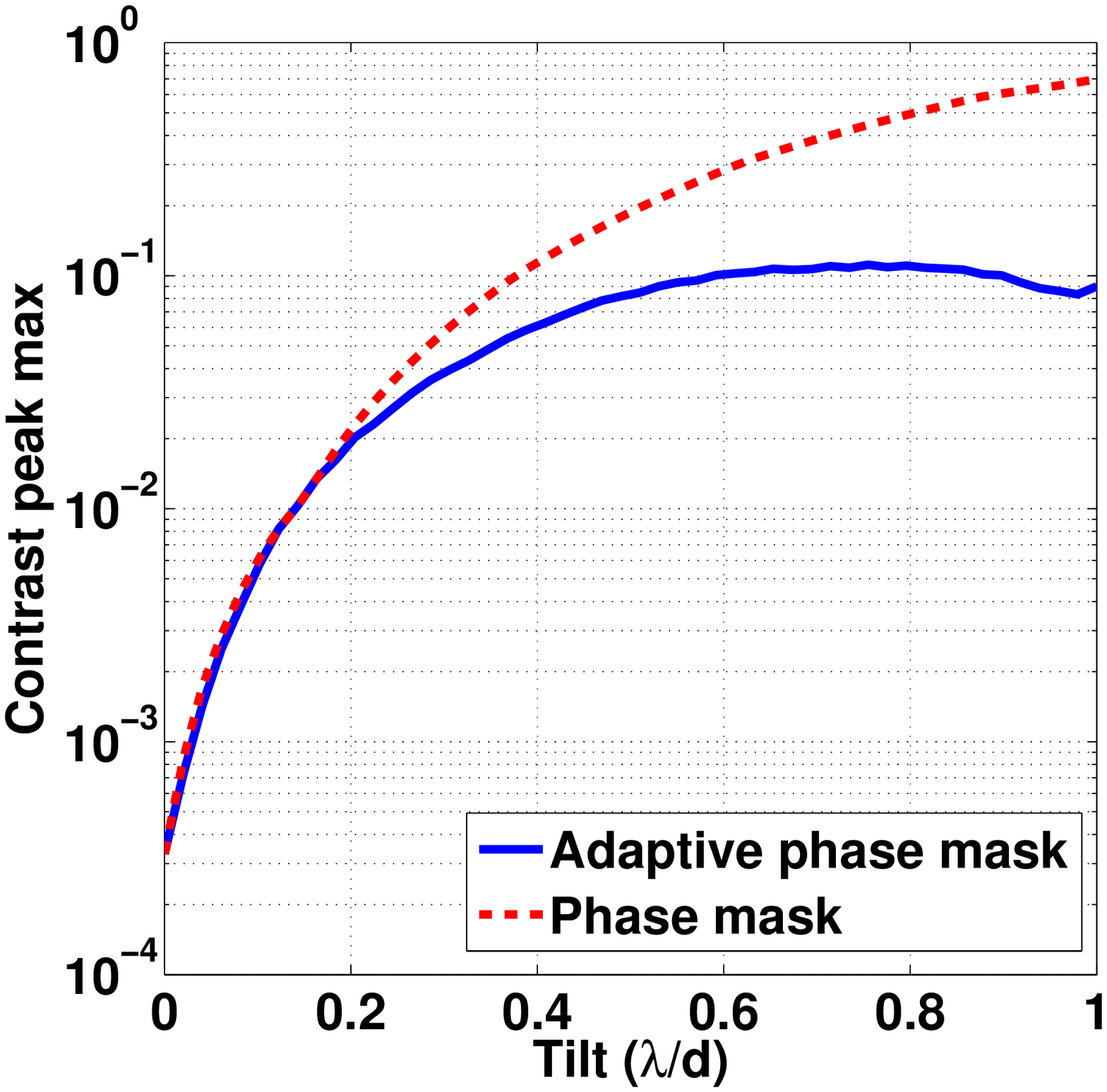}
\includegraphics[scale=0.4]{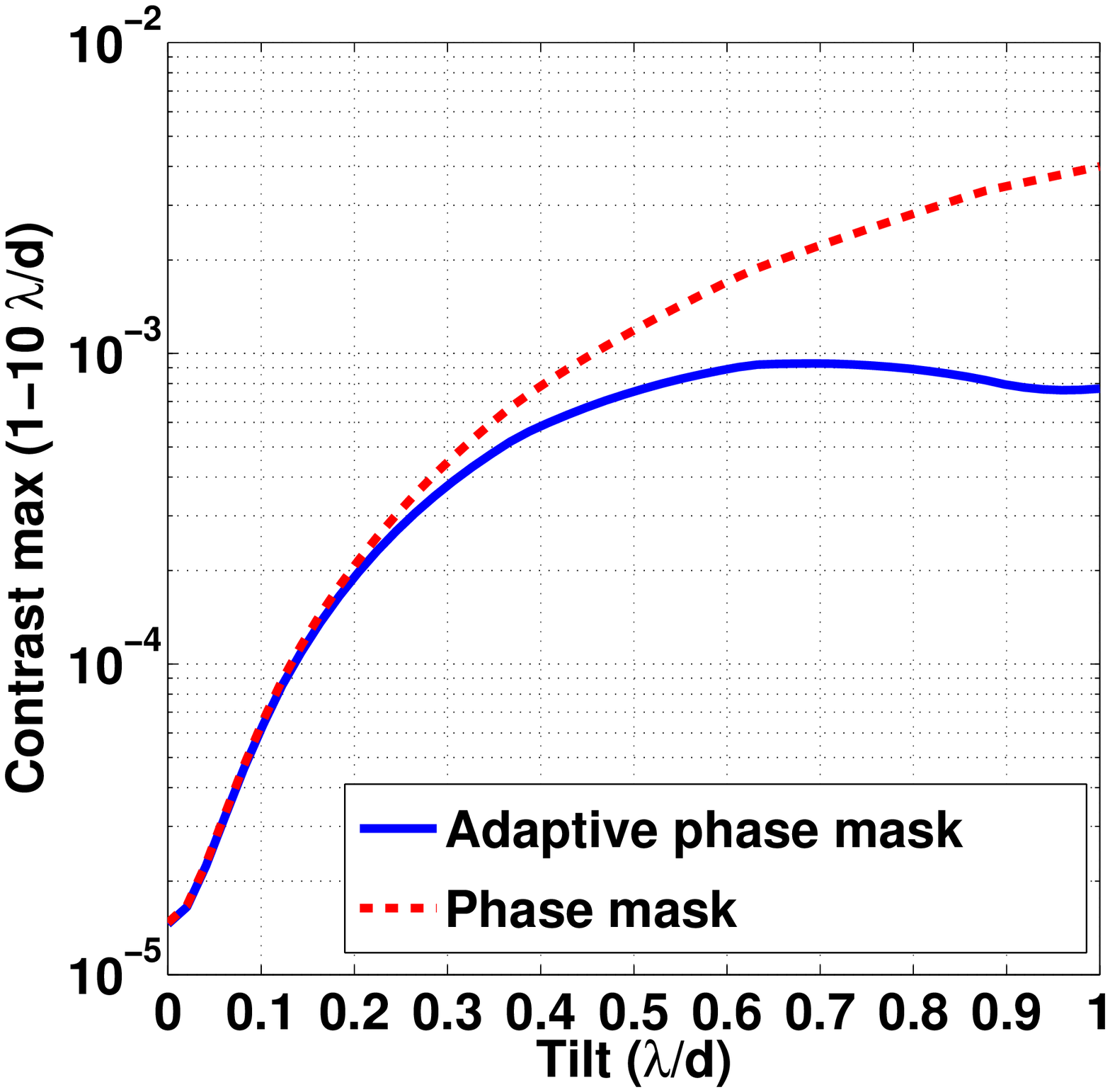}
  \caption{The curves represent the contrast obtained on the peak (left plot) or from the first to the tenth rings (right plot) for a perfect static phase mask (red curve) or for the adaptive phase mask (blue curve). The red curve represents the variation of the contrast without any control of transmission outside of the phase mask (it corresponds to the cut of top line of the Fig.~\ref{fig4}). The blue curve illustrates the contrast achieved by the control of transmission with an increasing tilt. The adaptive phase mask is more efficient over the rings where the contrast is needed and it maintains the contrast bellow $10^{-3}$ with a tilt of $\lambda/D$. It should be noted that the red curves are obtained for a purely achromatic phase mask whose diameter match perfectly the theoretical diameter balancing the amplitudes of the interfering waves. Any practical implementation in a real environment would show a much worse contrast especially for small tilts. \label{fig5}}
\end{figure*}

Fig.~\ref{fig4} shows how the tip-tilt degrades the nulling efficiency of the peak and over the first to the tenth rings (top line of each plots, also displayed as the red curves in Fig.~\ref{fig5} ) but also how adapting the transimission of the outer ``annulus'' region allows mitigating this effect (blue curves in Fig.~\ref{fig5}). The results displayed in Fig.~\ref{fig4} are obtained with an initial phase mask diameter slightly smaller than the perfect theoretical one balancing exactly the amplitudes of the two phase shifted waves. For that reason the optimal nulling is obtained for an outer transmission smaller than 1. 
In Fig.~\ref{fig5}, it should be noted that the red curve is a purely theoretical curve as no implementation of a static phase mask will ever be perfectly achromatic and match the exact diameter to balance the amplitudes of the interfering waves. In practice, the size of the phase mask combined to the chromatism will strongly degrade the initial contrast (at zero Tilt) of the red curve, nevertheless the control of transmission will keep the blue curve at the best efficiency allowed by the chromatism, the optimization of the occulting process will be then substantially improved by the adaptive phase mask. A potential upgrade of the adaptive mask to alleviate the imperfect overlap of the diffracted wave in the presence of tip-tilt would allow modulating the mask amplitude in a series of pie charts zones in the outer ``annulus'' electrode, associated with more resolution elements in the APM camera (Bourget et al.~2012, in preparation).

\section{Multi-Stage Adaptive Phase Mask}\label{sec:muti}
\begin{figure}[!ht]
  \centering
\includegraphics[scale=0.8]{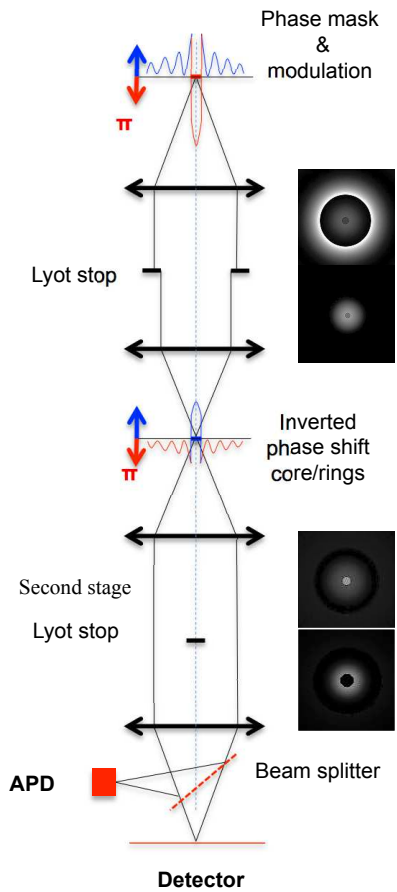}
  \caption{On the first adaptive phase mask, the phase shift is obtained by rotation of the polarization as shown in Fig.~\ref{fig2}. The control loop on the transmission outside of the phase mask is done at the first stage of the coronagraph. At the second stage, the adaptive phase mask applies the reversed rotation of the polarization by exchanging the orientation of the twist applied to the nematic liquid-crystal. On the right side of the figure,the pupils before and after the pupil Lyot stop are displayed for each stage. The diffraction effects on the masks are rejected outside of the pupil contour image at the first stage and inside the pupil contour image at the second stage. The remnant flux around the central obstruction image at the first stage is injected inside the central obstruction image at the second stage. The detection of the nulling by the APD is done at the output of the second stage. The second adaptive mask is passive, its effect on the destructive interference must be taken into account by the APD during the optimization of the nulling before closing the loop. \label{fig6}}
\end{figure}

In the case of a Lyot coronagraph, the occulting mask diffraction injects high frequencies inside the pupil image and especially on the border of the pupil. The Lyot diaphragm is used to stop that diffraction effect allowing high contrast imaging but at the cost of a loss of flux and resolution. For an on-axis telescope, the central obstruction introduces a diffraction effect that will also inject high frequencies inside the pupil image of the central obstruction; this contribution will easily be suppressed by a stop placed on the central obstruction image without loosing flux. For a phase mask coronagraph, the distribution of the high frequencies due to the diffraction on the mask is inverted compared to the Lyot coronagraph (First stage of Fig.~\ref{fig6}). High frequencies are rejected outside of the pupil image; a Lyot diaphragm stops them without loosing flux and resolution, the pupil size is unchanged. Nevertheless, a central obstruction will introduce high frequencies outside its pupil image then inside the useful pupil. As described by \citet{Haguenauer2005}, the contribution of the central obstruction will strongly affect the efficiency of the interference process. Using a Four-Quadrant Phase Mask \citet{Rouan2000}, numerical simulations shows that a 1.5-meter sub-pupil of the Palomar telescope (without central obstruction) allows reaching higher contrast than with the entire 5.1m pupil (with central obstruction). 
The Multi-Stage Vortex Coronagraph concept, developed by \citet{Mawet2011}, proposes a method to significantly attenuate the contribution of the diffraction induced by the central obstruction. Using the optical properties of the Vector Vortex Coronagraph \citep{Mawet2005}, the high frequencies injected in the pupil by the central obstruction can be removed by a stop without loosing light in a second stage of vortex coronagraphs. This action is allowed if the phase ramp of the two Vortex are inverted; in that case the high frequencies injected in the pupil by the central obstruction are transferred inside the central obstruction image in the second stage of coronagraph and then stopped by a mask. Following the Multi-Stage Vortex Coronagraph concept, the axisymmetric action of the two stages Vortex is applicable to other phase masks such as the four quadrants phase mask (FQPM) and the Roddier\&Roddier phase mask. The numerical simulation of a two-stage phase mask coronagraph with a $\pi$ phase shift inverted at each image plane was performed to verify the diffractions effects of the central obstruction for an on-axis telescope (Fig.~\ref{fig6}). Note that the second adaptive mask is passive. Its effect on the destructive interference must be taken into account by the second-stage APD during the optimization of the nulling before closing the loop.

\section{Conclusions and perspectives}
This paper presented the innovative concept of Adaptive Phase Mask, which was shown to allow an active optimization of the nulling process of a Roddier\&Roddier disk phase mask coronagraph. In this first paper, we focused on the active control of the effects of both the phase-mask diameter with respect to wavelength, and the image instabilities (tip-tilt, Strehl ratio variability, etc.). High band pass control is provided by the APD and Liquid-crystals properties. The chromatic effect of the phase shift is minimized using the geometrical/Pantcharatnam/Berry phase. The flexibility introduced by the geometric phase and the use of liquid crystals is thus very promising regarding the adaptability of the mask in both size and phase. Note that more complex phase and amplitude profiles could also be rendered to implement adaptable versions of the dual zone phase mask coronagraph \citep{NDiaye2012}.

In the case of an on-axis telescope, the central obstruction diffraction effect is strongly decreased by a two-stage coronagraph following the two-stage Vortex coronagraph concept developed by \citet{Mawet2011}. 

The flexibility offered by the use of the  geometrical/Pantcharatnam/Berry phase and dynamically driven liquid-crystals enables a whole new kind of coronagraph designs. As already mentioned, the segmentation of the outer region in pie charts to provide a mean of correction for contrast loss due to image motion is already being studied. Furthermore, the capability to modulate, in phase or in amplitude, localized areas of the image or pupil plane may be used to help discriminate between companion and speckles. Finally, the temporal modulation enabled by this technology could be associated with synchroneous detection techniques to help retrieve low contrast signals from a noisy quadi-static background.

\begin{acknowledgements}
The authors would like to thank the referee, Dr.~Laurent Pueyo for its very valuable remarks and comments, which helped greatly improve the quality and readability of this paper. This work was carried out at the European Southern Observatory (ESO) site of Vitacura (Santiago, Chile). 
\end{acknowledgements}

\bibliographystyle{aa} 
\bibliography{pbourget_2012_ref2_2012-10-21} 
\end{document}